# The impact of edges and dopants on the work function of graphene nanostructures. The way to high electronic emission from pure carbon medium.

D.G. Kvashnin[1], P.B. Sorokin[1,2], J. W. Brüning[3] and L. A. Chernozatonskii[1]

[1] Emanuel Institute of Biochemical Physics, 4 Kosigina Street, Moscow, Russia, 119334
[2] Technological Institute of Superhard and Novel Carbon Materials, 7a Centralnaya Street, Troitsk, Moscow, Russia, 142190
[3] Humboldt-Universität zu Berlin, Unter den Linden 6, Berlin, Germany 10099

*The impact of the edges and presence of dopants to the work function (WF) of graphene nanoribbons (GNR) and nanoflakes was studied by an ab initio approach. The strong dependence of the WF upon the GNR structure was found and a promising character for the field emission by the donor type impurities was observed. Basing on the predominant impact of the nanostructure edges to the emission properties, the small graphene flakes were investigated as a possible source for the electron emission. The obtained weak dependence of the low WF values of the graphene flakes on their size and shape allows to suggest that the pure carbon medium with high and uniform emission properties can be fabricated by today technology.*

Carbon-based nanostructures are a promising material for the application as a field electron emission source. For example, shortly after first identification in 1991[1] in Refs. 2-4, it was found out that carbon nanotubes are promising for cold electron emission. Despite of successful realization[5] and high efficiency of such nanostructures, the complicated fabrication of the carbon nanotube arrays of a uniform size hindered them from application in real devices. Graphene, which was synthesized only several years ago, is currently considered as a base for the whole future nanoelectronics. It is already applied as an element in the nanoelectronic schemes (high-frequency transistors,[6] logic transistiors[7]), as touch screens,[8] sensors,[7] supercapacitors,[9,10] and more.[7] Recently, an individual single-layer graphene has been considered as a source[11,12] for the field electron emission (FEE). In the case of a perfect graphene sheet, the value of the work function (WF, the main feature of the FEE effect) has been defined as 4.60 eV,[13] (in agreement with the theoretical data, 4.48 eV[14] LDA, 4.49 eV[15] GGA) which is a relatively high value. Therefore, the work function value decrease is highly desirable for successful graphene application in the FEE area. The work function reduction by 1 eV leads to an increase in the field emission current by over two orders of magnitude, which is suggested by the Fowler-Nordheim theory.[16] The WF of graphene can be controlled by the electric field effect (EFE), it was found [17] that the scanning Kelvin probe microscope application to the back-gated graphene devices allows to change the work function value within the 4.5 – 4.8 eV range for a single-layer graphene, and 4.65 – 4.75 eV for a bilayer graphene. The reference atoms introduction into the graphene lattice can significantly improve the FEE characteristics. The theory [18,19] suggests that the metal adatoms can dramatically decrease the WF of the graphene-based nanostructures, and that was confirmed by an experiment. [20] In the Ref. 21 it was found that the Cs doping of the five-layer carbon nanotubes leads to the work function drop from 4.8 eV to 3.8 eV, depending on the dopant concentration. In the case of single-wall nanotubes, the work function value decreased from 3.1 eV to 2.4 eV.[22]

The electrons emit mainly from the graphene edges, [23] and therefore their presence is highly desirable for an effective electron emission. The WF and field emission decreasing effect was observed for the ~1-10 nm graphene nanoflakes manufactured by the state-of-the-art technology. [24] The availability of the relatively simple fabrication methods for small graphene flakes with atomically precise shapes [25] allows to talk about the future application of such a material in the FEE field.

Also, the graphene edges passivation type is important for the FEE as well. In the study [26] was found out that the work function of a 14AGNR armchair graphene nanoribbon with different functional species equals to 4.1 eV for the hydrogenated edges, and 4.6 eV for the oxygen edge passivation.

In this study, an integrated research of the field emission properties of the graphene-based structures of different shapes was performed. The emission properties of the zigzag graphene nanoribbons (ZGNR) with edges passivated by different kinds of atoms and doped by various dopant atoms were studied using the *ab initio* methods. We didn't study here the energy stable reconstructed zigzag edges predicted early [27, 28, 29] due to the fact that we implied that graphene during the emission should be deposed on the metallic surface on which reczag edges are less stable than zigzag one. [30, 31] The dependence of the work function on the edge passivation types (H, F, clear edges) was studied, and the decreasing of WF value from 4.50 eV (clear edges) to 4.00 eV (H, F passivation) was observed. Also, it was found that there is an impact of the impurities to the ZGNR emission properties, in particularly, it was found that the nitrogen impurities with a less than 3 % concentration in ZGNR decrease the work function down to 3.0 eV. This result explains the observation of a low FEE in the nitrogen-doped graphene. [32] Finally, the dependence of the WF of the graphene flakes with hydrogenated edges upon the size and shape was studied. It was found that the emission properties of the flakes mainly depend on the perimeter (the length of the edges). The pronounce dependence of the work function for the small flakes was observed, whereas the WF values of the flakes with a perimeter larger than 3 nm are practically the same as the WF of the graphene nanoribbons with hydrogenated edges. The obtained results can be used for the design of the graphene-based materials with high emission properties.

The plane wave DFT PBE [33] electronic structure calculations of the carbon nanostructures were performed using the Ultrasoft Vanderbilt pseudopotentials [34] and a plane-wave energy cutoff of 30 Ry by a PWSCF code. [35] To calculate the equilibrium atomic structures of the graphene ribbons, the Brillouin zone was sampled according to the Monkhorst–Pack [36] scheme with a 1×1×24 k-point convergence grid, whereas only gamma-point was used for the graphene flakes relaxation. To avoid the interactions between the species, the neighboring structures were separated at least by 15 Å in the rectangular supercells.

The work function of the graphene nanoribbons was determined as the difference between the vacuum level and the Fermi level. The vacuum level was set to be an average electrostatic potential energy along the direction normal to each of the studied graphene-based nanostructure surface. In the case of semimetallic species, the Fermi level was defined as the highest occupying the quantum state in a system.

Firstly, let us consider the relatively narrow 4ZGNR and 8AGNR with a 0.71 nm and 0.83 nm width, respectively. We have studied the ribbons with the hydrogenated, fluorinated, and clean (without any passivation) edges.

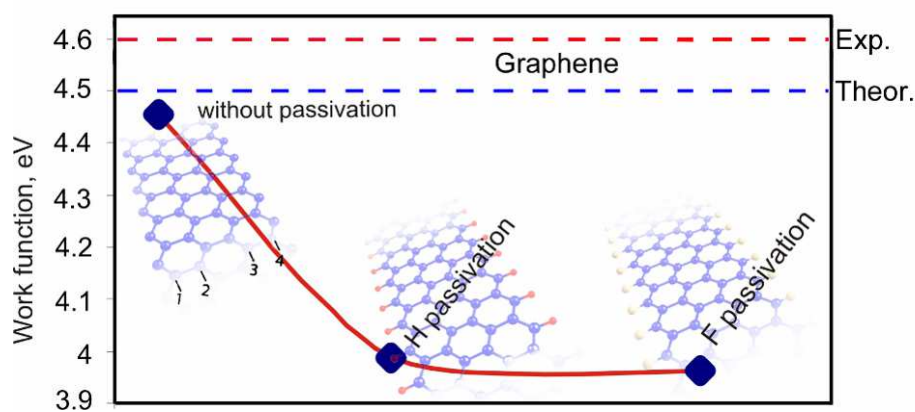

FIG. 1. Dependence of the 4ZGNR work function on the ribbon edges fictionalization. The atomic structures of the nanoribbon with a hydrogenated, fluorinated, and clean edges are shown. The red and blue dashed lines correspond to the theoretical (see Ref. 14 and 15) and experimental (see Ref. 13) values of the graphene work function.

The ribbons work function is sensitive to the passivation of the edges.[37] As it can be seen from the Fig. 1, the highest value of the work function (4.46 eV) corresponds to a ZGNR with clean edges, whereas the passivation reduces this value. In a case of hydrogen and fluorine passivation, the work function displays lower values: 3.98 eV and 3.96 eV, respectively. For the AGNR, slightly higher values were obtained: 3.88 eV, 3.75 eV and 4.87 eV for hydrogen, fluorine passivation and clean edges, respectively (the value 3.88 eV corresponds well with the reference data [26] which additionally justifies the chosen approach). For a further study, the graphene zigzag nanoribbons with hydrogen passivation were chosen.

We studied the impact of the doping to the emission properties of the nanoribbons. Nitrogen, boron, and phosphorus were chosen as dopants as the most natural for carbon doping elements. The concentration of doping atoms in the considered nanoribbons was about 3 % which is in the experimental range 1…10 % [32,38] for such type of doping for GNR. In the case of 4ZGNR, the ribbon's supercell contains one doping atom, 24 carbon atoms, and 6 hydrogen atoms (3.3 % of doping).

The work function on the doping type and doped atoms position in the nanoribbon structures was studied (Fig. 2). The boron-doped GNR displays the highest WF value among all considered cases, and it displays a strong dependence on the dopant position. Both phosphorous and nitrogen doping lead to the lowest WF value.

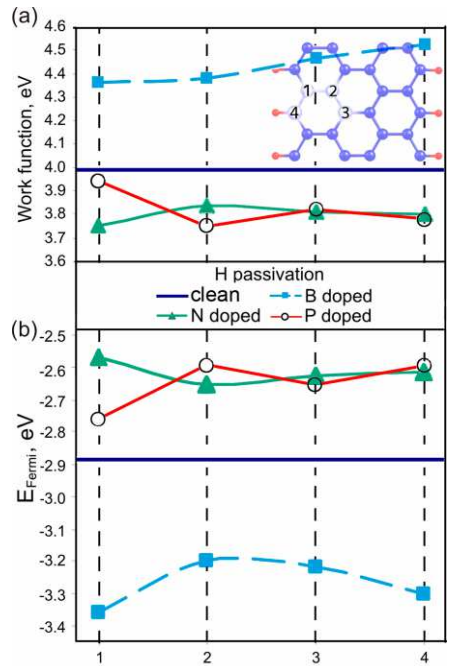

FIG. 2(Color online) Dependence of work function (a) and Fermi level (b) on the positions and type of the dopants. Only symmetry nonequivalent positions of the doping atoms were considered. The blue horizontal line corresponds to the work function (a) and Fermi level (b) of pure 4ZGNR.

The nitrogen, boron and phosphorous dopants strongly affect the Fermi level position which is directly related to the WF value. The donor-type doping leads to an increase in the Fermi level and the WF decrease, whereas an acceptor-type doping leads to the opposite behavior. For example, a nitrogen atom has one extra electron in comparison with a carbon atom. For an N-doped graphene ribbon, the extra electron energy level lies around the nitrogen atom and gives a rise to a donor state near the Fermi level (the similar results were obtained for the carbon nanotubes [39,40]). The impurity energy level depends on the Coulomb interaction of the extra electrons of the doping atoms and the π-edge unpaired electrons[41] Nitrogen doping induced a deep impurity level below the top of the valence band which induces an increase in the energy of the highest occupied level. An increase in the highest occupied level decreases the energy difference between the vacuum and Fermi levels, and therefore lowers the WF value.

The impurities tend to move out of the ribbons structure which is represented by a strain energy drop with a nitrogen atom approaching to the edge, see Fig. 3a (a similar result was obtained in Refs. 41, 42). But the work function does not significantly depend upon the dopant position and varies in the range from 3.76 to 3.84 eV.

For further considering of the impurities impact on the GNR work function value, a wider ribbon (6ZGNR) with a similar (to the previous case of 4ZGNR) concentration of the doping atoms was studied in detail. For this structure, we investigated eleven configurations of the various dopants positions (Fig. 3b). A 6ZGNR unit cell consists of 34 carbon atoms and 6 hydrogen atoms. The concentration of the doping atoms equals to 2.5%.

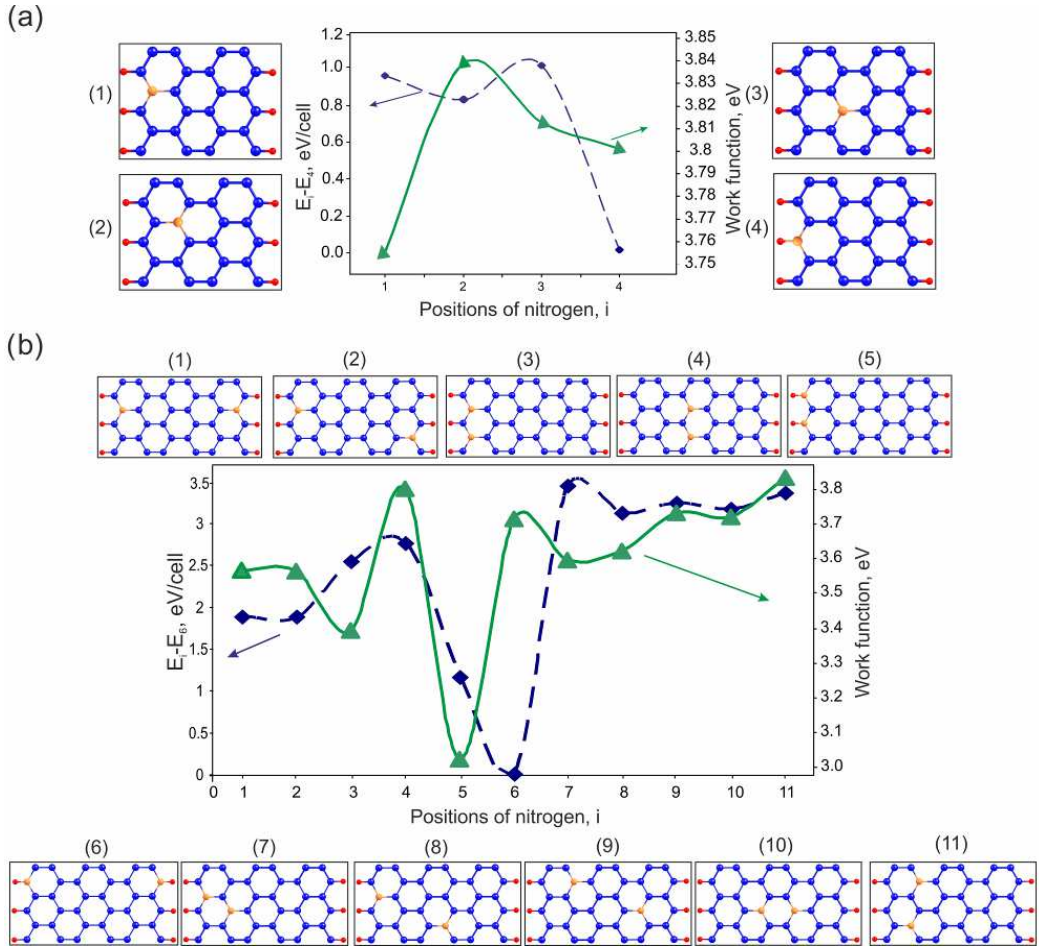

FIG. 3. (Color online) Dependencies of the strain energy and work function of the 4ZGNR (a) and 6ZGNR (b) with hydrogenated edges upon the nitrogen impurities in the ribbon structure. The strain energy is defined as a difference between the energy of the preferable position $E_4$ (a), $E_6$ (b), and the energy of current state, $E_i$. Only symmetry nonequivalent positions of the doping atoms were considered.

The configurations "1", "2", "3", "4" and "7" - "11" display the large strain energies (Fig 3b). The energy preferable is configurations with positions of the nitrogen impurities on the ribbon edge "5" and "6" which display the close strain energies, but drastically different WF values. The energy favorable is configuration "6" when two nitrogen atoms locate on the opposite sides, whereas the configuration "5" is less favorable due to the strain of the lattice on the edge induced by the presence of two neighbored nitrogen atoms. But the difference between the energies of configurations "5" and "6" is rather small (less than 0.03 eV/atom) which means that both configurations with WF values 3.0 and 3.7 eV, respectively, can be realized in nature.

The work function of both considered graphene ribbons with clean, undoped surface, does not display a clear dependence upon the width, and equals to ~ 4.0 eV. This fact suggests that the work function is mainly determined by the graphene edges and agrees with the previously reported suggestions. [23] Therefore, the most promising graphene nanostructures for the electron emission should be small graphene flakes due to the large contribution of the edges in their structure. The current experimental potential allows preparation of the nanometer-size [24] graphene flakes, and it is important to realize what size of the graphene flakes is most preferable for application in the FEE devices.

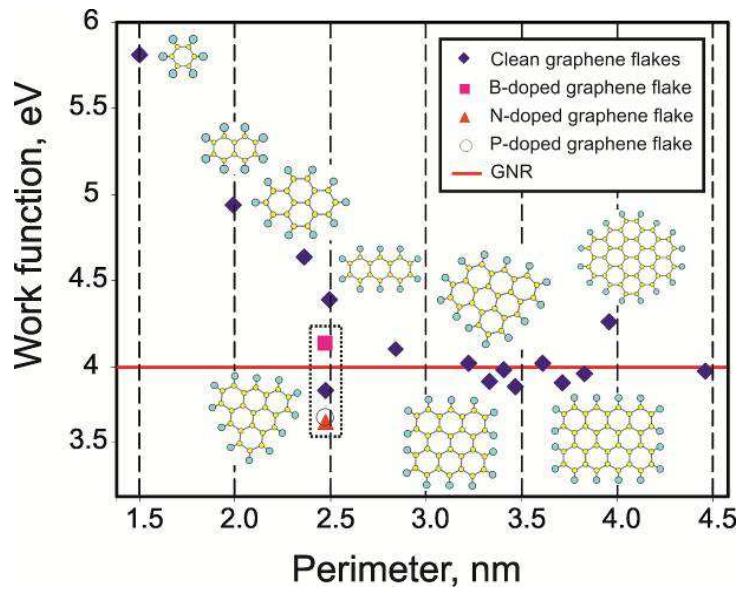

FIG. 4. (Color online) Dependence of the work function of the graphene nanoflakes on their perimeter.

Here we obtained the work function value for the graphene flakes with edges passivated by hydrogen of various shapes and sizes. It was found that WF depends mainly on the perimeter of the flakes edges (Fig. 4). The flakes with a perimeter larger than 3 nm display a nearly constant value of WF, close to the work function of the graphene nanoribbons (4 eV), whereas the smaller flakes display a much larger WF.

For clarity, we also investigated the work function of the graphene nanoflakes doped by nitrogen and boron. A structure with a 2.5 nm with a single dopant atom was studied (Fig. 4). Like in the case of the graphene nanoribbons, the presence of nitrogen and phosphorus atoms decreases the work function (from 3.8 eV to 3.6 eV), while the presence of a boron atom increases it (from 3.8 eV to 4.2 eV).

It is important to note that the WF of graphene nanostructures also can be modified by the interaction with various contacts and substrates. Such interaction can affect to the emission properties and will be investigated separately in the future work.

In summary, we have theoretically studied the impact of edges, impurities and size on the graphene work function. It was found that the zigzag graphene ribbons with hydrogen- and fluorine-passivated edges display the lowest WF values. We found out that the ZGNR doping of by boron, nitrogen, and phosphorous atoms (with a ~ 3% concentration) leads to a change in the work function value; the donor-type impurities (N and P) decrease the WF by ~ 0.3 eV. The dependence of the graphene flakes WF value on their size and shape was studied. It was found that the work functions of graphene flakes with a perimeter large than 3 nm are close to each other and nearly equal to the work function of the graphene ribbons (~ 4.0 eV). The doping of the graphene flakes by the donor-type impurities also decreases the WF. The weak dependence of the graphene flakes WF value upon their size and shape allows to conclude that the application of an experimental technique which permits cutting the graphene into small flakes with sizes larger than 3 nm should enable obtaining a medium with high and uniform emission properties.

This work was supported by the Russian Foundation for Basic Research (project no. 11-02-01453/12) and DFG 436 RUS 113/990/0-1. The authors are grateful to the Joint Supercomputer Center of the Russian Academy of Sciences and "Lomonosov" research computing center for the possibilities of using a cluster computer for the quantum-chemical calculations. D.G.K. acknowledges the support from the Russian Ministry of Education and Science (No. 948 from 21 of November 2012). P.B.S. also acknowledges the Russian Foundation for Basic Research (project no. 12-02-31261) and the Russian Ministry of Education and Science (Contract No. 14.B37.21.1645).